\documentclass[a4paper,11pt]{article}

\usepackage{pos}
\usepackage{siunitx}
\AtBeginDocument{\RenewCommandCopy\qty\SI}
\DeclareSIUnit\photon{\text{\ensuremath{\gamma}}}
\usepackage{hyperref}
\usepackage{subcaption}
\usepackage{wrapfig}
\usepackage{lineno}

\title{A co-deployed dust-logging instrument for the IceCube Upgrade and IceCube-Gen2}

\ShortTitle{A co-deployed dust-logging instrument}

\author{The IceCube Collaboration \\{\normalsize \normalfont(a complete list of authors can be found at the end of the proceedings)}\\}

\emailAdd{anna.eimer@fau.de}
\emailAdd{martin.rongen@fau.de}

\abstract{

A precise understanding of the optical properties of the instrumented Antarctic ice sheet is
crucial to the performance of optical Cherenkov telescopes such as the IceCube Neutrino
Observatory and its planned successor, IceCube-Gen2. One complication arising from the large envisioned footprint of IceCube-Gen2 is the larger impact of the so-called ice tilt. It describes the undulation of ice layers of constant optical properties within the detector.

In this contribution, we will describe the project to build a co-deployed laser dust logger. This is a device to measure the stratigraphy of impurities in the ice to derive the ice tilt. It consists of a light source that will be co-deployed with the photosensor modules, meaning it is part of the deployment string and operated during the deployment of the detector. The newly developed device will be tested during the deployment of the IceCube Upgrade in the 2025/26 austral summer to pave the way for IceCube-Gen2.

\vspace{4mm}

{\bfseries Corresponding authors:}
Anna Eimer$^{1*}$, 
Martin Rongen$^{1}$\\ 
{$^{1}$ \itshape Erlangen Centre for Astroparticle Physics (ECAP), Friedrich-Alexander-Universität Erlangen-Nürnberg}\\[4mm]
$^*$ Presenter
}

\ConferenceLogo{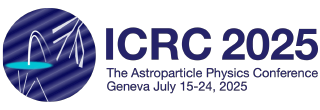}

\FullConference{39th International Cosmic Ray Conference (ICRC2025)\\
 15–24 July 2025\\
Geneva, Switzerland\\}

\begin{document}

\maketitle

\section{Ice stratigraphy}\label{ice_stratigraphy}


The IceCube Neutrino Observatory \cite{IceCube} is a cubic kilometer array of photosensors in the ice at the South Pole, detecting neutrinos. Neutrino interactions produce charged secondary particles that produce Cherenkov light. Since this Cherenkov light might get absorbed or scattered on its way from its origin in the ice to the detector module, it is crucial to know the optical properties of the ice for the reconstruction of neutrino properties from raw data. 

Due to the ice formation over thousands of years and different dust concentrations in the air at different times, ice with similar optical properties is arranged in layers \cite{journal_of_glaciology_2013}. These layers are deposited at the same time and thus have the same optical properties. Therefore, they are called isochrones. Looking at the isochrones over the whole depth of the ice gives the `ice stratigraphy'. Certain features make it possible to relate the ice at a certain depth to known geological events in time, e.g. volcanic eruptions \cite{journal_of_glaciology_2013} (visible as very narrow highly scattering layers) or global features like the `dust layer' with a high scattering coefficient over a larger depth range \cite{journal_of_glaciology_2013}.

\begin{figure}[b]
    \centering
    \includegraphics[width=\linewidth]{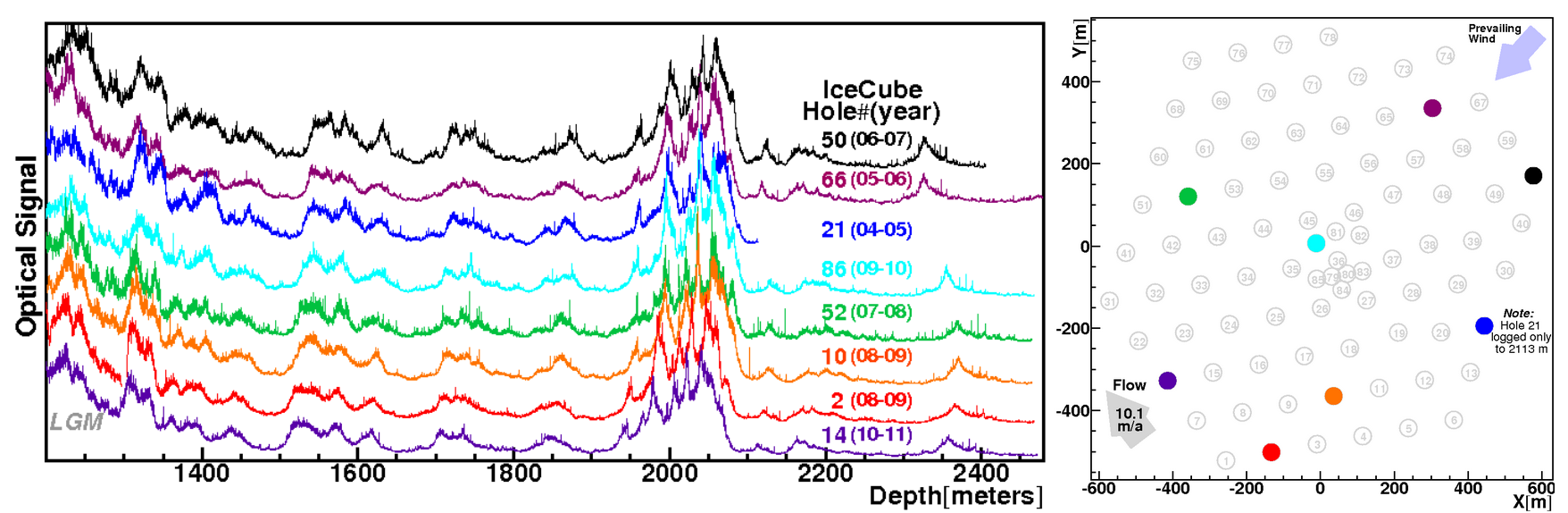}
    \caption{Stratigraphy measurements performed for IceCube. The left plot shows the optical signal in arbitrary units over the depth in meter. The colors in this plot correspond to the color coded positions in the detector layout on the right. (Taken from \cite{journal_of_glaciology_2013})}
    \label{fig:dust_logger_stratigraphy}
\end{figure}

One way of measuring the scattering of light in ice is the `laser-based dust logging' \cite{journal_of_glaciology_2013}. This device shines laser light horizontally into the ice. The light will then be scattered in the ice layer and a fraction scatters back to the instrument. An integrated photomultiplier tube (PMT) finally collects the scattered light. This principle is used over the whole depth of the IceCube detector. The result of these measurements for IceCube are shown in \autoref{fig:dust_logger_stratigraphy}. The signal strength is proportional to the amount of impurities in the ice. 

Additionally, these layers of optical properties are not flat over the full detector volume. The layers follow the shape of the bedrock beneath the glacier and flatten out towards the surface. 
These undulations, called `tilt' in IceCube, can be measured by logging the stratigraphy at different position within the instrumented ice. This was already done for IceCube with eight laser-based dust-logger measurements \cite{journal_of_glaciology_2013}. Those measurements needed a separate deployment immediately after drilling the hole, as well as a retrieval before the photosensors can be lowered. Therefore, this procedure was very time-consuming. Since then, several other methods were explored to measure the tilt. The currently used method in the IceCube ice model reconstructs the tilt with flasher LEDs incorporated in the IceCube digital optical modules (short DOMs) \cite{LEDtilt:ICRC2023}. Most recently, camera data was used for a feasibility study to measure the ice stratigraphy \cite{CameraLogging:2025icrc}.

\begin{wrapfigure}{r}{0.25\linewidth}
    \vspace{-1.2cm}
    \centering    
    \includegraphics[width=0.65\linewidth]{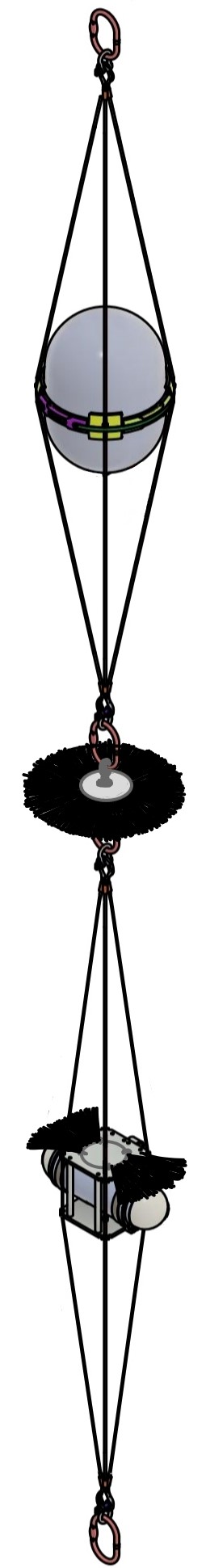}
    \caption{Sketch of the LOMlogger setup with the LOM at the top, a baffle halfway between the two modules and the POCAM at the bottom with another baffle attached on top of its harness.}
    \label{fig:enter-label}
    \vspace{-1.5cm}
\end{wrapfigure}

The tilt measurement will be crucial for the larger footprint and larger module spacing envisioned for IceCube-Gen2 \cite{Gen2paper}. Hence, a more practical version of the `dust logger', called the LOMlogger, will be tested in the low energy extension of IceCube deployed this year, called the IceCube Upgrade \cite{UpgradeProceedings}. This is done to show the feasibility for IceCube-Gen2.

\section{Working principle of the co-deployed laser dust-logger}\label{working_principle}

The LOMlogger needs, similar to the `dust logger', an emitting light source and a receiving photon detector. In the case of the LOMlogger, a modified POCAM \cite{Henningsen_2020} and the lowest detector module on the instrumentation cable are used. We call this concept co-deployed. Using parts that will have a purpose beyond the stratigraphy measurement on the instrumentation cable makes the operation more cost efficient and less time consuming.

The emitted light fans out horizontally in a \SI{60}{\degree} opening angle into the ice, until it gets scattered by dust particles. The extent in horizontal direction leads to more statistics for the measurement. A low beam divergence in vertical direction determines the resolution of the stratigraphy measurement. The used laser will be operated in a pulsed mode with an LD driver circuit, explained in \cite{Henningsen_2020}. The laser light source will be the main focus of the following proceeding.

After the first scattering process, it scatters randomly through the ice and eventually is detected by one of the PMTs in the detector module \SI{2}{m} above the emitter. In the case of the IceCube Upgrade version, the light is detected by a LOM \cite{LOM:2025icrc}. The LOM consists of \num{16} PMTs, which improves the directional reconstruction of the detected photons. The recorded data will be stored per PMT in \SI{6}{ms} time bins, which corresponds to one data point every \SI{1}{mm} depth.

In order to shield the LOM from light that scatters through the water column in the deployment borehole, baffles are needed. The black nylon brushes should absorb the light, but let the water pass through it. There are two positions for the baffles: one directly on top of the light source to prevent the light from entering the borehole as soon as possible and one halfway between the two modules to mitigate signal contribution from reflected rays. Another mitigation strategy uses the directionality of LOM PMTs. The PMTs facing the opposite side of the light source will likely collect only scattered photons coming through the ice, whereas the PMTs facing the light source might also collect photons reflected through the borehole.

\pagebreak

\section{Requirements for the optics}\label{requirements}

The LOMlogger project can learn from the dust logger measurements about its required capabilities. The dust logger data was reanalyzed similar to the approach reported in \cite{paleowind}. The data is also publicly available \cite{dust_map}. Using the dynamic time warping algorithm \cite{dtw}, the different stratigraphy measurements could be compared. Features as described in \autoref{ice_stratigraphy} could be matched even though appearing at different depths in the different measurements.


It is important to identify which features are needed to still be able to match two measurements. Therefore, the goal is to see which resolution is needed to recreate the original analysis of the depth difference performed by Bay et al. \cite{dust_map} within $\pm \SI{1}{m}$. 
To determine the resolution needed for a reasonable stratigraphy, the available dust logger data (one data point every \SI{2}{mm}) could be averaged to have one data point every \SI{20}{cm} before the dynamic time warping algorithm is not able to match features properly. This leads to the conclusion that features with a size of approximately \SI{20}{cm} are used by the algorithm to match different measurement. Hence, these features need to be visible in the stratigraphy measurements.


\begin{figure}[h]
    \centering
    \includegraphics[width=0.8\linewidth]{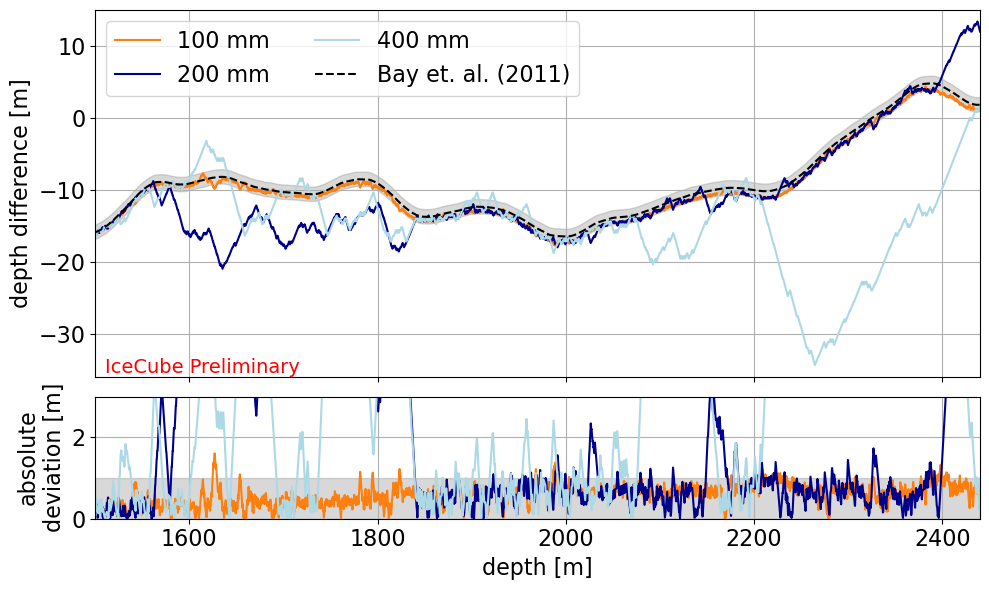}
    \caption{Alignment of two stratigraphy measurements over depth with different resolution (see labels in the legend). The black line represents the original analysis \cite{dust_map} and the gray band (representing $\pm \SI{1}{m}$) around it is the still allowed region for the fit to diverge from. The lower panel shows the absolute deviation between Bay et al. and this analysis with the different resolutions.}
    \label{fig:depth_resolution}
\end{figure}

\begin{figure}
    \centering
    \begin{subfigure}{0.45\linewidth}
        \centering
        \includegraphics[width=0.9\linewidth]{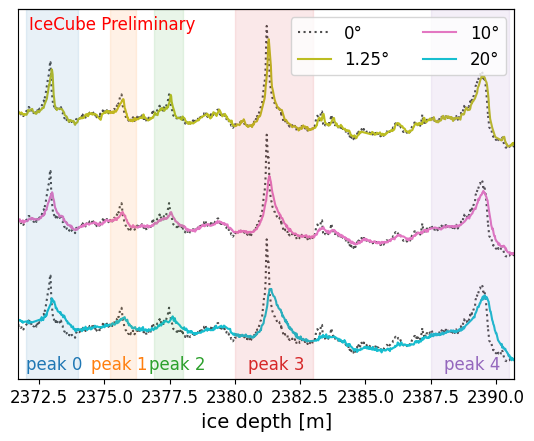}
        \caption{Exemplary stratigraphy simulation with different divergence angle settings for representative features (shaded regions).}
    \end{subfigure}
    \hspace{0.05\linewidth}
    \begin{subfigure}{0.45\linewidth}
        \centering
        \includegraphics[width=0.95\linewidth]{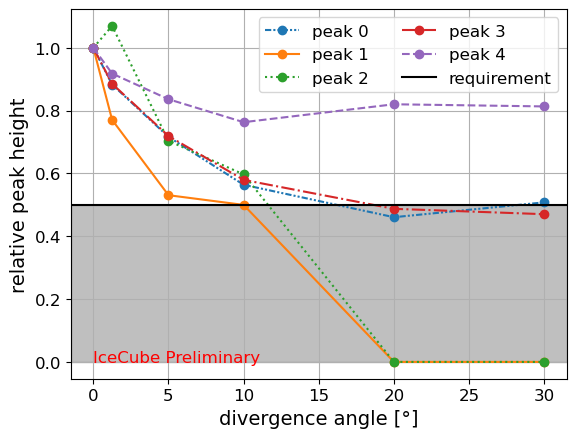}
        \caption{Changing divergence angle (x axis) degrades feature visibility shown as relative peak height on the y axis.}
    \end{subfigure}
    \caption{Simulation of degraded optical systems in a region with \num{5} representative \SI{20}{cm} features.}
    \label{fig:requirements}
\end{figure}

This result was used to find a representative region in the stratigraphy with \num{5} features of a FWHM of \SI{20}{cm} (called peaks in \autoref{fig:requirements}). Using the PPC simulation tool \cite{PPC}, the LOMlogger was simulated varying the vertical inclination and the beam divergence. The simulation delivers the anticipated stratigraphy where it uses the light source parameters as input. 
The beam divergence was changed to see when those features would degrade to \SI{50}{\percent} of the ideal peak height in the stratigraphy simulation. This resulted in a beam divergence that should be less than \SI{10}{\degree}.
Changing the vertical inclination of the beam, shifts the scanned depth area. Hence, the beam inclination should not surpass \SI{10}{\degree}.
These results lead to the incentive to make the beam divergence as small as possible and try to stabilize the device as much as possible during the deployment. The beam divergence will be discussed later on. The device stability during deployment is mostly out of our hands.

The other important parameter for this operation is the intensity of the laser. It should deliver enough photons that scattered photons reach the receiving LOM, but not enough to saturate it.





The fraction of emitted photons expected to be detected by each PMT was gauged from a PPC \cite{PPC} photon propagation simulation of the LOMlogger system. In the clearest ice, where the return probability is the lowest, it is \num{4.3e-8} per LOM PMT. The intensity ratio between the cleanest and the most scattering ice as observed by the old dust logger is 30. So the detection probability per LOM PMT in the most scattering ice is expected to be \num{1.3e-6}. The LOM PMTs start to saturate at $\sim$\SI{1000}{PE} per pulse, assuming a repetition frequency of \SI{1}{kHz}. So the highest permissible brightness is \num{8e8} photons per pulse. To still detect at least 10 photons per pulse in the cleanest ice the light source shall not be dimmer than \num{3e8} photons.



\section{Optics design}\label{optics}

\begin{figure}
    \centering
    \includegraphics[width=0.9\linewidth]{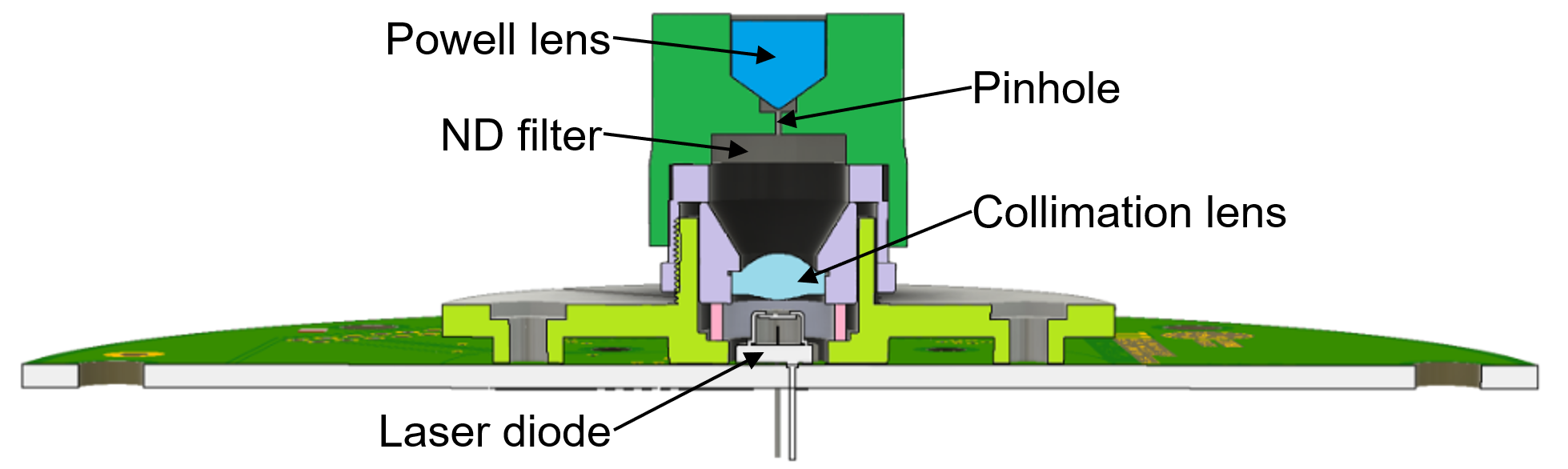}
    \caption{CAD model of the LOMlogger optics on the POCAM analog board (bottom part). Sitting directly on the board is the laser (white component) and the 3D-printed collimator holding structure (light green). The 3D-printed part contains parts of the Thorlabs Collimation Tube (spring depicted in pink, lens in light blue and collimation cap in violet). On top sits the 3D-printed Powell lens holding structure (dark green) with integrated cavity for the neutral density filter (gray rectangle), pinhole (small gray bottle neck) and Powell lens (dark blue part).}
    \label{fig:optics_system}
\end{figure}

The requirements lead to a few components for the optics system. The complete optics system, including the board that houses the laser is shown in \autoref{fig:optics_system}. In the following the different parts get described from the bottom (near the PCB) up.

As a light source a \SI{405}{nm} laser diode (RTL405-600MGE from Roithner) is used. The wavelength was chosen, because ice is the most transparent for \SI{405}{nm} light \cite{optical_properties_of_south_pole_ice}. This makes it the most efficient way to transport the light to the detector. The laser diode in particular was chosen, because it was already tested for a similar purpose \cite{laser}.

\begin{figure}[h]
    \centering
    \begin{subfigure}{0.45\linewidth}
        \centering
        \includegraphics[width=\linewidth]{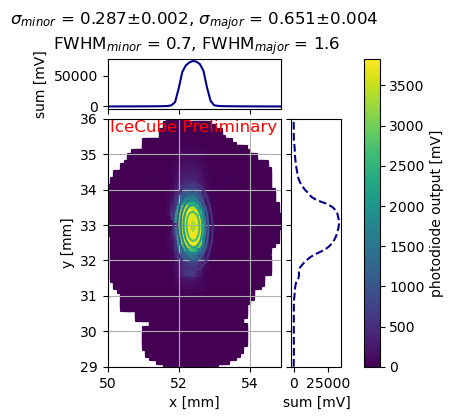}
        \caption{Collimated laser beam.}
    \end{subfigure}
    \hspace{0.05\linewidth}
    \begin{subfigure}{0.45\linewidth}
        \centering
        \includegraphics[width=\linewidth]{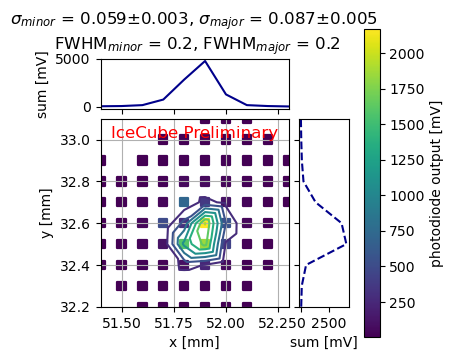}
        \caption{Collimated laser beam with pinhole.}
    \end{subfigure}
    \caption{2D-scan of the laser beam profile (lower left panel in the respective plots) with results for FWHM calculation and 2D-Gaussian fit (given on the top). All quantities are given in \si{mm}. The right panel in each plot shows the 1D-intensity profile in y-direction and the panel at the top in x-direction.}
    \label{fig:beam_profile_wo_powell}
\end{figure}

The laser beam will be focused by a Thorlabs LTN330-A Adjustable Laser Diode Collimation Tube. The collimation is needed to make the laser as convergent as possible. Testing the prototype this produced an ellipsoid beam profile with FWHM of \SI{0.7}{mm} and \SI{1.6}{mm} in x and y direction. This beam profile will be further restricted with a pinhole after the collimator to get the needed beam diameter of \SI{0.4}{mm} for the beam optics. Beam profiles of the collimated beam and after the pinhole can be found in \autoref{fig:beam_profile_wo_powell}. The 2D-scans are performed by a photodiode on a 2 dimensional linear stage. The FWHM was calculated using the one dimensional profile of the beam, by summing up the photodiode output over each column or row. Additionally, a 2D-Gaussian fit was performed as a cross check to get a handle on the beam size.

\begin{figure}[h]
    \centering
    \includegraphics[width=0.9\linewidth]{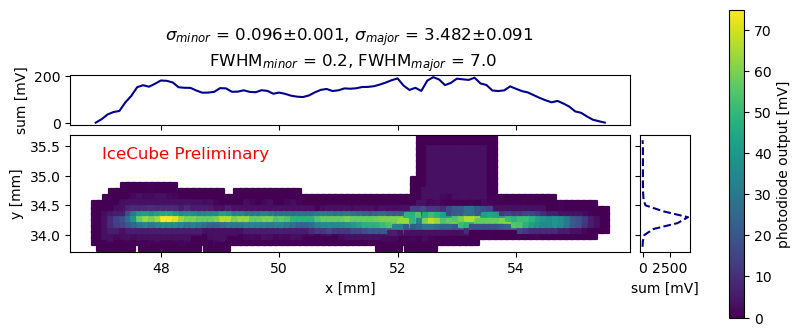}
    \caption{2D-scan of the laser beam profile after the Powell lens was added (lower left panel) with results for FWHM calculation and 2D-Gaussian fit (given on the top). All quantities are given in \si{mm}. The right panel in the plot shows the 1D-intensity profile in y-direction and the panel at the top in x-direction.}
\label{fig:beam_profile_powell}
\end{figure}

The requirements showed that a very thin beam in vertical direction is needed. On the other hand the more light shines into the ice in the horizontal direction the more ice in one layer can be probed. Therefore the beam needs to be shaped like a fan. This can be achieved by a LGL160 Powell lens. To produce a even distribution in intensity, the incoming beam needs to have a diameter smaller than \SI{0.4}{mm}. Hence, the pinhole is placed in front of the lens. A 2D-scan of the final beam going into the ice is shown in \autoref{fig:beam_profile_powell}.

\begin{figure}
    \centering
\includegraphics[width=0.589\linewidth]{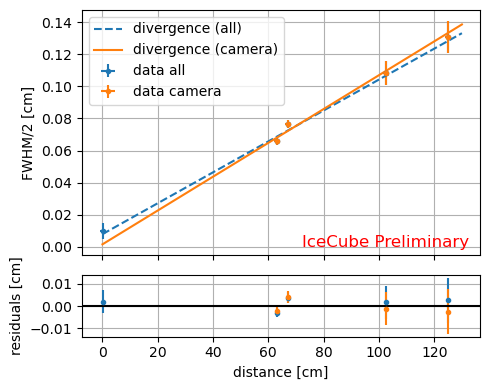}
    \caption{Result of the beam divergence measurement. Distance between laser and screen is plotted on the x-axis and half the FWHM is plotted in the upper panel. This is used to perform the linear fit leading to the divergence. The lower panel shows the residuals. The orange dots describe the data taken by the camera method, the blue dot describes the linear stage method.}
    \label{fig:divergence_result}
\end{figure}

Besides the overall beam profile the divergence of the beam with the Powell lens is important. This was tested with camera pictures taken at different distances between laser and detection screen (fluorescent paper). The divergence angle in y-direction can be calculated using the small angle approximation of the tangent. In this case, a linear fit of distance between laser and screen and half of the FWHM was performed. The result, the slope of the fit, is a divergence angle of \SI{0.06\pm0.01}{\degree} as shown in \autoref{fig:divergence_result}. As can be seen in the plot the camera method and the linear stage method described above deliver compatible results.

The laser intensity can be reduced with a neutral density filter, if needed. This measure accounts for laser diode intensity variations. The filter can be placed between collimator and pinhole. The reduction of the intensity by a factor of \num{4} was sufficient in most cases.

\section{Conclusion}

The ice stratigraphy is an important way to improve the neutrino event reconstruction in IceCube. It is the basis for the current ice model and can be improved by further measurements conducted during the deployment of the IceCube Upgrade. It is also important regarding the tilt expected in IceCube-Gen2's larger footprint.

The stratigraphy in the IceCube Upgrade can be measured with a newly developed version of the dust logger using a laser light source (modified POCAM) shining light into the ice while lowering the instrumentation cable into the ice. The light scattered in the ice is detected by one of the photosensor modules (LOM). The two instruments are currently in production and will be deployed this austral summer.

In order to measure the stratigraphy certain requirements for the optics need to be fulfilled. The vertical beam divergence should be less than \SI{10}{\degree} and also the inclination of the beam should be less than \SI{10}{\degree} to get a resolution smaller than \SI{20}{cm}. Additionally, the intensity should be within the range of \SIrange{3e8}{8e8}{\photon} per pulse to surpass the background, but not saturate the LOM.

These requirements lead to a beam design that is very narrow in the vertical direction, but can be broad in the horizontal direction. This can be achieved with a \SI{405}{nm} laser diode collimated and shaped by a pinhole to an appropriate size for the Powell lens. The Powell lens produces a fan of light with the desired dimensions. 

\bibliographystyle{ICRC}
\bibliography{references}

%

\clearpage

\section*{Full Author List: IceCube Collaboration}

\scriptsize
\noindent
R. Abbasi$^{16}$,
M. Ackermann$^{63}$,
J. Adams$^{17}$,
S. K. Agarwalla$^{39,\: {\rm a}}$,
J. A. Aguilar$^{10}$,
M. Ahlers$^{21}$,
J.M. Alameddine$^{22}$,
S. Ali$^{35}$,
N. M. Amin$^{43}$,
K. Andeen$^{41}$
% ...
\\
\\
% ...
$^{10}$ Universit{\'e} Libre de Bruxelles, Science Faculty CP230, B-1050 Brussels, Belgium \\
% ...
$^{16}$ Department of Physics, Loyola University Chicago, Chicago, IL 60660, USA \\
$^{17}$ Dept. of Physics and Astronomy, University of Canterbury, Private Bag 4800, Christchurch, New Zealand \\
% ...
$^{21}$ Niels Bohr Institute, University of Copenhagen, DK-2100 Copenhagen, Denmark \\
$^{22}$ Dept. of Physics, TU Dortmund University, D-44221 Dortmund, Germany \\
% ...
$^{35}$ Dept. of Physics and Astronomy, University of Kansas, Lawrence, KS 66045, USA \\
% ...
$^{39}$ Dept. of Physics and Wisconsin IceCube Particle Astrophysics Center, University of Wisconsin{\textemdash}Madison, Madison, WI 53706, USA \\
% ...
$^{41}$ Department of Physics, Marquette University, Milwaukee, WI 53201, USA \\
% ...
$^{43}$ Bartol Research Institute and Dept. of Physics and Astronomy, University of Delaware, Newark, DE 19716, USA \\
% ...
$^{63}$ Deutsches Elektronen-Synchrotron DESY, Platanenallee 6, D-15738 Zeuthen, Germany \\
$^{\rm a}$ also at Institute of Physics, Sachivalaya Marg, Sainik School Post, Bhubaneswar 751005, India \\
% ...

\subsection*{Acknowledgments}

\noindent
The authors gratefully acknowledge the support from the following agencies and institutions:
USA {\textendash} U.S. National Science Foundation-Office of Polar Programs,
U.S. National Science Foundation-Physics Division,
U.S. National Science Foundation-EPSCoR,
U.S. National Science Foundation-Office of Advanced Cyberinfrastructure,
Wisconsin Alumni Research Foundation,
Center for High Throughput Computing (CHTC) at the University of Wisconsin{\textendash}Madison,
Open Science Grid (OSG),
Partnership to Advance Throughput Computing (PATh),
Advanced Cyberinfrastructure Coordination Ecosystem: Services {\&} Support (ACCESS),
Frontera and Ranch computing project at the Texas Advanced Computing Center,
U.S. Department of Energy-National Energy Research Scientific Computing Center,
Particle astrophysics research computing center at the University of Maryland,
Institute for Cyber-Enabled Research at Michigan State University,
Astroparticle physics computational facility at Marquette University,
NVIDIA Corporation,
and Google Cloud Platform;
Belgium {\textendash} Funds for Scientific Research (FRS-FNRS and FWO),
FWO Odysseus and Big Science programmes,
and Belgian Federal Science Policy Office (Belspo);
Germany {\textendash} Bundesministerium f{\"u}r Forschung, Technologie und Raumfahrt (BMFTR),
Deutsche Forschungsgemeinschaft (DFG),
Helmholtz Alliance for Astroparticle Physics (HAP),
Initiative and Networking Fund of the Helmholtz Association,
Deutsches Elektronen Synchrotron (DESY),
and High Performance Computing cluster of the RWTH Aachen;
Sweden {\textendash} Swedish Research Council,
Swedish Polar Research Secretariat,
Swedish National Infrastructure for Computing (SNIC),
and Knut and Alice Wallenberg Foundation;
European Union {\textendash} EGI Advanced Computing for research;
Australia {\textendash} Australian Research Council;
Canada {\textendash} Natural Sciences and Engineering Research Council of Canada,
Calcul Qu{\'e}bec, Compute Ontario, Canada Foundation for Innovation, WestGrid, and Digital Research Alliance of Canada;
Denmark {\textendash} Villum Fonden, Carlsberg Foundation, and European Commission;
New Zealand {\textendash} Marsden Fund;
Japan {\textendash} Japan Society for Promotion of Science (JSPS)
and Institute for Global Prominent Research (IGPR) of Chiba University;
Korea {\textendash} National Research Foundation of Korea (NRF);
Switzerland {\textendash} Swiss National Science Foundation (SNSF).

\end{document}